\documentstyle[aps,psfig,multicol]{revtex}
\newcommand{\bleq}{\ifpreprintsty
                   \else
                   \end{multicols}\vspace*{-3.5ex}{\tiny
                   \noindent\begin{tabular}[t]{c|}
                   \parbox{0.493\hsize}{~} \\ \hline \end{tabular}}
                   \fi}
\newcommand{\eleq}{\ifpreprintsty
                   \else
                   {\tiny\hspace*{\fill}\begin{tabular}[t]{|c}\hline
                    \parbox{0.49\hsize}{~} \\
                    \end{tabular}}\vspace*{-2.5ex}\begin{multicols}{2}
                    \fi}
\newcommand{\bcols}{\ifpreprintsty\else\begin{multicols}{2}\fi}
\newcommand{\ecols}{\ifpreprintsty\else\end{multicols}\fi}

\newcommand{\bq}{\begin{equation}}
\newcommand{\ee}{\end{equation}}

\newcommand{\eps}{\varepsilon}

\begin{document}
\draft
\title{Scaling at the chaos threshold in an interacting quantum dot}
\author{X. Leyronas$^{1}$\footnote{Present address:
Laboratoire de Physique Statistique, Ecole Normale Sup\'erieure,
24 rue Lhomond, 75231 Paris Cedex 05, France.
}, P. G. Silvestrov$^{1,2}$, and C. W. J. Beenakker$^{1}$}
\address{$^1$ Instituut-Lorentz, Universiteit Leiden,
P.O. Box 9506, 2300 RA Leiden, The Netherlands\\
$^2$ Budker Institute of Nuclear Physics, 630090 Novosibirsk, Russia
}
\date{November, 1999}
\maketitle
\begin{abstract}
The chaotic mixing by random two-body interactions of many-electron Fock
states in a confined geometry is investigated numerically and compared
with analytical predictions. Two distinct regimes are found in the
dependence of the inverse participation ratio in Fock space $I$ on the
dimensionless conductance of the quantum dot $g$ and the excitation energy
$\varepsilon$. In both regimes $I\gg 1$, but only the small-$g$ regime
is described by the golden rule. The crossover region is characterized
by a maximum in a scaling function that becomes more pronounced with
increasing excitation energy.  The scaling parameter that governs the
transition is $(\varepsilon/g)\ln g$.
\end{abstract}
\bcols

The highly excited atomic nucleus was the first example of a quantum
chaotic system, although the interpretation of Wigner's distribution of
level spacings \cite{Wig67} as a signature of quantum chaos came many
years later, from the study of electron billiards \cite{Boh84}. While
the spectral statistics of the nucleus and the billiard is basically
the same, the origin of the chaotic behavior is entirely different
\cite{Guh98}: In the billiard chaos appears in the single-particle
spectrum as a result of boundary scattering, while in the nucleus chaos
appears in the many-particle spectrum as a result of
interactions.

The study of the interaction-induced transition to chaos entered
condensed matter physics with the realization that a semiconductor
quantum dot could be seen as an artificial atom or compound nucleus
\cite{Mar96}. A particularly influential paper by Altshuler, Gefen,
Kamenev, and Levitov \cite{Alt97} studied the interaction-induced decay
of a quasiparticle in a quantum dot and interpreted the broadening of
the peaks in the single-particle density of states as a
delocalization transition in Fock space.
Different scenario's leading to a smooth rather than an abrupt
transition from localized to extended states were considered
later~\cite{Jac97,Mir97,Sil97}. Recent computer simulations
\cite{Mej98,Ley99} also confirm the smooth crossover from
localized to delocalized regime for quasiparticle decay.

As emphasized by Altshuler et al. \cite{Alt97}, the delocalized
regime in the quasiparticle decay problem is not yet chaotic
because the states do not extend uniformly over the Fock space.
One may study the transition to chaos in the single-particle density
of states, but theoretically it is easier
to consider instead the mixing by interactions of {\em
arbitrary\/} many-particle states. This was the approach taken
in Refs.\
\cite{Jac97,Sil97,Fla97,Geo98,Ber98,Sil98}, focusing on two
quantities:
The distribution of the energy level spacings and the inverse
participation ratio (IPR) of the wave functions in Fock space.
Both quantities can serve as a signature for chaotic behavior,
the spacing distribution by comparing with Wigner's distribution
\cite{Wig67} and the IPR by comparing with the golden rule
(according to which the IPR is the mean spacing $\delta$ of the
many-particle states divided by the mean decay rate $\Gamma$ of
a non-interacting many-particle state \cite{Geo98}).  Two
fundamental questions in these investigations are: 1) What is
the scaling parameter that governs the transition to chaos? 2)
How sharp is the transition?

In a recent paper \cite{Sil98} one of us presented analytical
arguments for a singular threshold governed by the scaling
parameter $x=(\varepsilon/g\Delta)\ln g$, where $\Delta$ is the
single-particle level spacing, $\varepsilon$ is the excitation
energy, and $g$ is the conductance in units of $e^{2}/h$. (Both
$\varepsilon/\Delta$ and $g$ are assumed to be $\gg 1$.) In
contrast, Georgeot and Shepelyansky \cite{Geo98} argued for a
smooth crossover governed by the parameter
$y=(\varepsilon/g\Delta)\sqrt{\varepsilon/\Delta}$. (The same
scaling parameter was used in  Refs.\ \cite{Jac97,Ber98}.) The
parameter $y$ is the ratio of the strength $V\sim \Delta/g$ of
the screened Coulomb interaction
\cite{Alt97,Bla96} and the energy spacing $\Delta_2 \sim
(\varepsilon/\Delta)^{-3/2}\Delta$ of states that are directly
coupled by the two-body interaction \cite{Jac97}.  The parameter
$x$ follows if one considers contributions to the IPR that
involve the effective interaction of $2,3,4,\ldots$ particles.
Subsequent terms in this series are smaller by a factor  $(\ln
g/g)\Delta_{n}/\Delta_{n+1}$, where $\Delta_{n} \sim
(\varepsilon/\Delta)^{-n+1/2}\Delta$ is the spacing of states
that are coupled by an effective interaction of $n$ particles
\cite{Sil98}.  (The large logarithm $\ln g$ appears in the
expansion parameter because of the large contribution from
intermediate states whose energies are close to the states to be
mixed.)

The purpose of this paper is to investigate the
interaction-induced transition to chaos by exact
diagonalization of a model Hamiltonian.
We concentrate on the IPR because
for that quantity an analytical prediction exists
\cite{Sil98} for the $\varepsilon$ and $g$-dependence. (There is
no such prediction for the spacing distribution.) The numerical
data is consistent with a chaos threshold at a value of $x$ of
order unity.

The model for interacting spinless fermions that we study is the layer
model introduced in  Ref.\ \cite {Geo98} and used for the quasiparticle
decay problem in Ref.\ \cite{Ley99}.
The Hamiltonian is $H=H_0 +H_1$, with
\begin{eqnarray}\label{Ham}
H_0=\sum_{j} \varepsilon_j c^{\dagger}_j
c^{\vphantom{\dagger}}_j,\qquad
H_1=\sum_{i<j,k<l} V_{ij,kl}c^{\dagger}_l c^{\dagger}_k
c^{\vphantom{\dagger}}_i c^{\vphantom{\dagger}}_j .
\end{eqnarray}
The single-particle levels $\varepsilon_j$ are uniformly
distributed in the interval
$[(j-\frac{1}{2})\Delta,(j+\frac{1}{2})\Delta]$. The interaction
matrix elements $V_{ij,kl}$ are zero unless $i,j,k,l$ are four
distinct indices with $i+j=k+l$. The (real) non-zero matrix
elements have a Gaussian distribution with zero mean and
variance $V^2 =(\Delta/g)^2$. The Fock states are eigenstates
of $H_0$, given by Slater determinants of the occupied levels
$k_1,k_2,k_3,\ldots$. The interaction mixes Fock states for
which $\sum_{p} k_{p}$ equals a given integer. (Without this
restriction the model is the same as the two-body random-interaction
model introduced in nuclear physics \cite{Fre70,Boh71}.)
The excitation
energies of the states with given $k_1,k_2,k_3,\ldots$ lie in a
relatively narrow layer (width of order $j^{1/4}\Delta$) around
the mean excitation energy $j\Delta$.  The number of states
in the $j$-th layer is the number of partitions ${\cal P}(j)$ of
$j$. For our largest $j=26$ this number is ${\cal
P}(26)=2436$, which is still tractable for an exact
diagonalization. Without the decoupling of the entire Fock space
into distinct layers, such large excitation energies would not
be accessible numerically. The layer approximation becomes
more reasonable for larger $g$, because then $V\ll\Delta$
so that states from different layers may be regarded as uncoupled.

The inverse participation ratio
$I=\sum_{m}|\langle\alpha|m\rangle|^4$ of the eigenstate
$|\alpha\rangle$ of $H$ is the inverse of the number of
eigenstates $|m\rangle$ of $H_0$ that have significant overlap
with $|\alpha\rangle$. We calculate $I$ as a function of $g$
for different layers $j$, corresponding to a mean excitation
energy $\varepsilon=j\Delta$. The IPR fluctuates strongly from state
to state and for different realisations of the random matrix $H$. We
calculate the averages $\overline{I}$, $\overline{1/I}$ and $\overline{\ln
I}$ where the overline $\overline{\cdots}$ indicates an average
both over the ${\cal P}(j)$ states $|\alpha\rangle$ in the
$j$-th layer and over some $10^{3}$ realisations of $H$. We
first consider the logarithmic average $\overline{\ln I}$, for
which the fluctuations are smallest.

\begin{figure}
\centerline{\psfig{figure=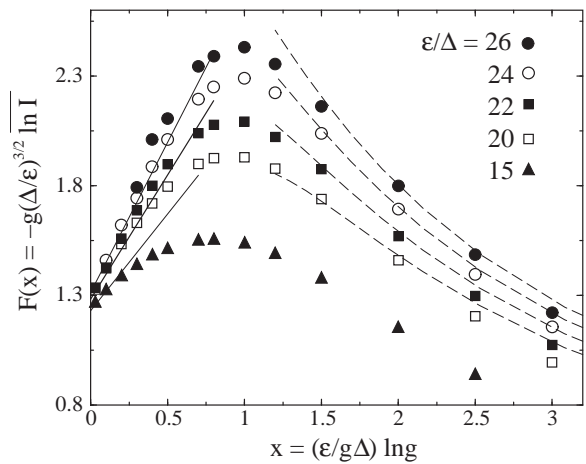,width= 8cm}}
\medskip
\caption[]{
Average logarithm of the inverse participation ratio $I$ as
a function of the dimensionless conductance $g$, in rescaled
variables. The different sets of data points follow from the
layer model for different excitation energies
$j=\varepsilon/\Delta$. Statistical errors are smaller than
the size of the markers. The straight solid lines are the
analytical prediction (\ref{eqFn}) of the scaling theory,
without any adjustable parameters.  (Only the lines for
$\varepsilon/\Delta = 15, 20 $, and $26$ are shown for clarity.)
The
dashed curves are the golden-rule prediction (\ref{Igr}), with a
single adjustable parameter (the same for all curves, but the data
for $\varepsilon/\Delta=15$ was left out of the fit).
\label{fig1}}
\end{figure}

In Fig.\ \ref{fig1} we have plotted the numerical data for the
$g$-dependence of $\overline{\ln I}$, for different values of
$\varepsilon/\Delta$. In order to compare with the analytical prediction
of Ref.\ \cite{Sil98}, we have rescaled the variables such that Fig.\
\ref{fig1} becomes a plot of $- y^{-1} \overline{\ln I}$ versus $x$.
The prediction is that, in the thermodynamic limit \cite{thermodynamic},
the scaling function $F(x) = - y^{-1} \overline{\ln I}$ depends
only on $x$ for $x\lesssim 1$. This scaling behavior can not be
checked directly because finite-size effects introduce an additional
$\varepsilon$-dependence into the function $F(x)$. This is why we can
not directly test whether $x$ or $y$ is the correct scaling parameter.
Fortunately, it is possible to include finite-size effects into the
scaling function and test the theory in this way.

Applying the method of Ref.\ \cite{Sil98} for the calculation of
$\overline{\ln I}$ one finds that the function $F(x)$
in the thermodynamic limit has the Taylor
series
\begin{equation}
\label{eqF}
F(x) = - y^{-1}
\overline{\ln I}=\sum_{n=0}^{\infty} c_n
x^n ,
\end{equation}
with corrections of order $1/\ln g$. All coefficients $c_{n}$ are
positive. The scaling behavior (\ref{eqF}) is expected to be
universal (valid for any model with random two-body
interactions), but the coefficients $c_n$
are model specific. The first two coefficients for the layer model are
\begin{equation}
\label{c01}
c_0=\frac{8(2-\sqrt{2})}{\sqrt{3\pi}}= 1.53 \ \ , \ \ c_1=
\frac{81}{25}\sqrt{\frac{2}{\pi}}\, c_0= 3.95 \ .
\end{equation}
Finite-size effects introduce an $\varepsilon$-dependence
into the coefficients, through the $\varepsilon$-dependence of
$K_n \equiv (\Delta/\Delta_n)(\Delta/\varepsilon)^{n-1/2}$.
The series expansion of $F(x)$ in terms of the $K_n$'s is
\begin{eqnarray}
\label{eqFn}
 F(x) = 4(\sqrt{2}-1)\sqrt{\pi}K_2 + 36(2-\sqrt{2})K_3 x +{\cal
O}(x^2).
\end{eqnarray}
For $\varepsilon/\Delta \rightarrow \infty$ we have
$K_2 \rightarrow (2/\pi)\sqrt{2/3}=0.5198$,
$K_3 \rightarrow 6\sqrt{6}/25\pi=0.1871$ and we recover the
thermodynamic limit (\ref{eqF}).  For the excitation energies
$\varepsilon/\Delta =15,20,22,24,26$ of the simulation, one has
$K_2=0.419,0.436,0.439,0.444,0.447$ and
$K_3=0.0414,0.0536,0.0577,0.0615,0.0648$.  The resulting small-$x$
behavior of the scaling function is plotted in Fig.\ \ref{fig1}
(solid lines) and agrees quite well with the numerical data.

Analytically, the scaling function $F(x)$ is only known for
$x\ll 1$. In the simulation, we observe a maximum of $- y^{-1}
\overline{\ln I}$ at $x\simeq 1$. The maximum becomes more
pronounced with increasing excitation energy. We argue that it
is a signature of the transition to chaos, because beyond the
maximum, for $x\gtrsim 1$, the IPR is observed to follow the
golden-rule prediction (see discussion below)
\begin{eqnarray}
\label{Igr}
I_{\rm{golden-rule}} = C [j^{5/4}{\cal P}(j)]^{-1}g^2 .
\end{eqnarray}
This golden-rule prediction is shown dashed in Fig.\ \ref{fig1},
with the coefficient $C\approx 0.51$ as the single fit
parameter. (The smallest $\varepsilon/\Delta=15$ was left out of the
fit.)
Note that $-y^{-1}\ln I_{\rm{golden-rule}}$ has
a maximum for an IPR of order unity, hence in the regime of
localized states. In contrast, the maximum in
$-y^{-1}\overline{\ln I}$ occurs
when the IPR is $\ll 1$, hence in the regime of extended states.
We now discuss the small and large-$x$ regimes in some more detail.

The large-$x$ regime is described by the golden rule $I_{\rm
golden-rule}=\delta/\Gamma$, according to which all basis states
within the decay width $\Gamma$ of a non-interacting state are
equally mixed into the exact eigenstate. This complete mixing
amounts to fully developed chaos. For our model the level
spacing of the many-particle states in the $j$-th layer is
$\delta\sim j^{1/4}\Delta/{\cal P}(j)$ and the Breit-Wigner
width is $\Gamma\sim V^{2}/\Delta_{2}\sim j^{3/2}g^{-2}\Delta$,
which leads to Eq.\ (\ref{Igr}).  One notices in Fig.\
\ref{fig1} that for the largest $x$ the data points fall
somewhat below the golden-rule prediction. This is due to the
finite band width of the layer model.  The IPR saturates at
$3/{\cal P}(j)$ \cite{Mej98}
 when the decay width $\Gamma$ becomes comparable to
the band width $j^{1/4}\Delta$. The corresponding upper bound on
$x$ for the validity of the golden rule is $x\lesssim j^{3/8}\ln
g$.  The finite band width of the layer model becomes less
significant for large $j$, which is why the agreement with the
golden rule improves with increasing $j$.

The small-$x$ regime is described by the scaling function $F(x)$.
The term of order $x^{n}$ in the Taylor series (\ref{eqF}) contains the
$n+1$-th order effective interaction $V^{\rm eff}_{n+1}$ between $n+2$
particles and holes.  A Fock state in the $j$-th layer contains about
$\sqrt{j}$ excited particles and holes \cite{Fermi}.  Because this is
a large number for $j\gg 1$, the IPR factorizes into a product
of independent
contributions from $2,3,4,\ldots$ interacting particles,
\bq\label{Veff}
\overline{\ln I}\sim \sum_{n=0}^{\infty}
\overline{|V_{n+1}^{\rm eff}|}/\Delta_{n+2}.
\ee
A calculation of $\overline{|V_{n+1}^{\rm eff}|}$ leads to Eq.\
(\ref{eqF}). The appearance of the modulus of the matrix element
in Eq.\ (\ref{Veff}) is easily understood for the case of only
two unperturbed many-particle states interacting via the matrix
element $V^{\rm eff}$. The IPR changes by order unity if two
Fock states come energetically within a separation $|V^{\rm eff}|$ of each
other. The probability of such a near degeneracy is small like
$|V^{\rm eff}|/\Delta$. (There is no level repulsion for the
many-particle solutions of the non-interacting Hamiltonian.) Because
for weak interaction the IPR can change significantly but only
with a small probability, the IPR fluctuates strongly. Indeed,
in our simulations much larger statistics was necessary in order
to reach good accuracy in the small-$x$ regime. (The remaining
statistical error in Fig.\ \ref{fig1} is smaller than the size
of the markers.)

\begin{figure}
\centerline{\psfig{figure=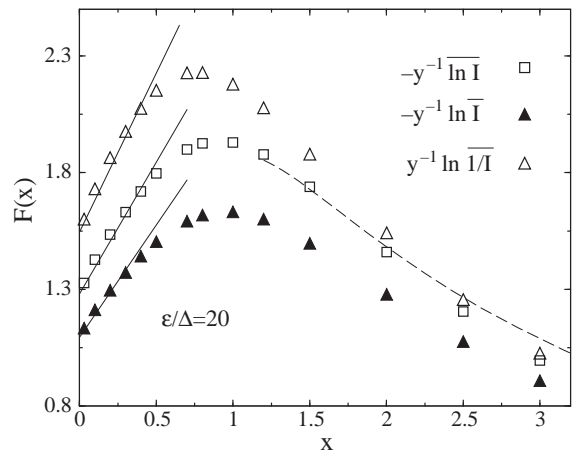,width= 8cm}}
\medskip
\caption[]{
Averages $-\overline{\ln I}, -\ln \overline{I}$, and $\ln
\overline{1/I}$ as a function of $g$, rescaled in the same way
as in Fig.\ \ref{fig1}, for $\varepsilon/\Delta=20$.  For small
$x$, the three averages follow the scaling theory
(\ref{threeav}) (solid lines).
For large $x$ the averages $-\overline{\ln I}$ and $\ln
\overline{1/I}$ follow the golden rule (dashed line).
\label{fig2}}
\end{figure}

In Fig.\ \ref{fig2} we compare the logarithmic average
$\overline{\ln I}$ with
the two other averages $\ln\overline{I}$ and $-\ln\overline{1/I}$.
Within the small-$x$ regime of validity of Eq.\ (\ref{eqF})
the three averages are related by
\begin{equation}\label{threeav}
\overline{\ln I}=2(2-\sqrt{2})\ln
\overline{I}=-2(\sqrt{2}-1)\ln \overline{1/I}.\label{averages}
\end{equation}
These numerical coefficients do not depend on the number of
particles involved in the interaction (cf.\ the explicit calculation of
$\overline{I}$ in Ref.\ \cite{Sil98}). As one can see in Fig.\ \ref{fig2},
for $\varepsilon/\Delta=20$, the relation (\ref{threeav}) agrees
well with the simulation. In the chaotic regime, for large $x$, Eq.\
(\ref{threeav}) is no longer valid. The average  $-\ln\overline{1/I}$,
which is dominated by the majority of states having a large
number of components, is
close to $\overline{\ln I}$ at large $x$. The average $\ln\overline{I}$
is dominated by rare states with an anomalously small number of components
and falls below the two other averages. This indicates an asymmetric
distribution of $\ln I$ in the chaotic regime for the layer model.

So far we have only addressed the question of the scaling
variable that governs the transition to chaos. What remains is
the question: How sharp is the transition? The singular
threshold predicted in Ref.\ \cite{Sil98} developes only in the
thermodynamic limit and would be smoothed by finite-size effects
in any simulation. The corresponding non-analyticity of
$\overline{\ln I}$ is related to the high-order behavior of the
series (\ref{eqF}). Since our numerics allows to distinguish
only the first two coefficients $c_0$ and $c_1$, it leaves open
the question about the non-analyticity. Still, even if the
series (\ref{eqF}) would be absolutely convergent, the resulting
smooth function of the single variable $x$ could not describe
the IPR for large $x$ because it is incompatible with the golden
rule $-y^{-1}{\ln I_{\rm golden-rule}}\sim x^{-1}\ln g$.  This
different scaling behavior for small and large values of $x$
suggests that the peak observed in Fig.\ \ref{fig2} would evolve
into a singular threshold in the thermodynamic limit. The only
way to maintain a smooth crossover would be to introduce a
parametrically large interpolating region between the two
different scaling regimes. We can not exclude this interpolating
region on the basis of the numerical data, however,
theoretically \cite{Sil98} there is no indication for such a
region.

In summary, by exact diagonalization of a model Hamiltonian we
have presented evidence for an interaction-induced transition to
chaos in a quantum dot. Upon inclusion of finite-size effects, a
good agreement is obtained  with the scaling theory of Ref.\
\cite{Sil98}, supporting the assertion that
$x=(\varepsilon/g\Delta)\ln g$ is the scaling parameter for the
transition. The different behavior of the
scaling function for small and large $x$ suggests that the
transition would become a singular threshold in the
thermodynamic limit.

This work was supported by the Dutch Science Foundation NWO/FOM
and by the TMR program of the European Commission. The research of
PGS was supported by RFBR, grant 98-02-17905. Discussions with
J.~Tworzyd\l o are gratefully acknowledged.

\ecols

\begin{references}
\bibitem{Wig67} E. P. Wigner, SIAM Rev.\ {\bf 9}, 1 (1967).
\bibitem{Boh84} O. Bohigas, M.-J. Giannoni, and C. Schmit, Phys.\ Rev.\
Lett.\ {\bf 52}, 1 (1984).
\bibitem{Guh98} T. Guhr, A. M\"{u}ller-Groeling, and
H. A. Weidenm\"{u}ller, Phys.\ Rep.\ {\bf 299}, 189 (1998).
\bibitem{Mar96} Th.\ Martin, G. Montambaux, and J. Tr\^an Thanh V\^an,
editors, {\em Correlated Fermions and Transport in Mesoscopic Systems}
(Editions Fronti\`eres, Gif-sur-Yvette, 1996).
\bibitem{Alt97} B. L. Altshuler, Y. Gefen, A. Kamenev, and L. S. Levitov,
Phys.\ Rev.\ Lett.\ {\bf 78}, 2803 (1997).
\bibitem{Jac97} P. Jacquod and D. L. Shepelyansky, Phys.\ Rev.\ Lett.\
{\bf 79}, 1837 (1997).
\bibitem{Mir97} A. D. Mirlin and Y. V. Fyodorov, Phys.\ Rev.\ B {\bf 56},
13393 (1997).
\bibitem{Sil97} P. G. Silvestrov, Phys.\ Rev.\ Lett.\ {\bf 79}, 3994
(1997).
\bibitem{Mej98} C. Mej\'\i a-Monasterio, J. Richert, T. Rupp, and
H. A. Weidenm\"uller, Phys.\ Rev.\ Lett.\ {\bf 81}, 5189 (1998).
\bibitem{Ley99} X. Leyronas, J. Tworzyd{\l}o, and C. W. J. Beenakker,
Phys.\ Rev.\ Lett.\ {\bf 82}, 4894 (1999).
\bibitem{Fla97} V. V. Flambaum and F. M. Izrailev, Phys.\ Rev.\
E\ {\bf 56}, 5144 (1997).
\bibitem{Geo98} B. Georgeot and D. L. Shepelyansky, Phys.\ Rev.\ Lett.\
{\bf 79}, 4365 (1998).
\bibitem{Ber98} R. Berkovits and Y. Avishai, Phys.\ Rev.\ Lett.\
{\bf 80}, 568 (1998).
\bibitem{Sil98} P. G. Silvestrov, Phys.\ Rev.\ E {\bf 58}, 5629 (1998).
\bibitem{Bla96}Ya. M. Blanter, Phys.\ Rev.\ B\ {\bf 54}, 12807 (1996).
\bibitem{Fre70} J. B. French and S. S. M. Wong, Phys.\ Lett.\ B {\bf 33},
449 (1970); {\bf 35}, 5 (1971).
\bibitem{Boh71} O. Bohigas and J. Flores, Phys.\ Lett.\ B {\bf 34}, 261 (1971);
{\bf 35}, 383 (1971).
\bibitem{thermodynamic} By the thermodynamic limit we mean the
limit of high excitation energy and large conductance, namely,
$\eps,g\rightarrow\infty$ at fixed $x=(\eps/\Delta)(\ln g/g)$.
\bibitem{Fermi} The probability to find the single-particle level $\eps_j$
(with excitation energy $j\Delta$) occupied in a Fock state
is $p(j)=(e^{j/T}+1)^{-1}$. It has the form of a Fermi-Dirac
distribution at
an effective temperature $T=\sqrt{6j}/\pi$. This formula, though formally
valid only in the thermodynamic limit $T\gg 1$, describes
well the average occupation number already at the values
$T=3.5-4$ of our simulations.

\end{references}
\end{document}